\documentclass[prd,twocolumn,superscriptaddress,longbibliography,preprintnumbers,floatfix,nofootinbib]{revtex4-1}

\usepackage{url}
\usepackage{xspace}
\usepackage{dsfont}
\usepackage{amssymb}
\usepackage{amsmath}
\usepackage{graphicx}
\usepackage{comment}
\usepackage[caption=false]{subfig}
\usepackage[colorlinks=true,citecolor=blue]{hyperref}
\usepackage{multirow}
\usepackage{hhline}

\newcommand{\flow}{DLE~} 
\newcommand{\classi}{LRE~}

\begin{document} 

\title{Discriminative versus Generative Approaches to Simulation-based Inference}

\author{Benjamin Sluijter}
\email{benjaminsluijter@gmail.com}
\affiliation{Leiden Institute of Physics, Universiteit Leiden, 2300 RA Leiden, 
The Netherlands}
\affiliation{Physics Division, Lawrence Berkeley National Laboratory, Berkeley, CA 94720, USA}

\author{Sascha Diefenbacher}
\email{sdiefenbacher@lbl.gov}
\affiliation{Physics Division, Lawrence Berkeley National Laboratory, Berkeley, CA 94720, USA}

\author{Wahid Bhimji}
\email{wbhimji@lbl.gov}
\affiliation{National Energy Research Scientific Computing Center, Berkeley Lab, Berkeley, CA 94720, USA}

\author{Benjamin Nachman}
\email{bpnachman@lbl.gov}
\affiliation{Physics Division, Lawrence Berkeley National Laboratory, Berkeley, CA 94720, USA}
\affiliation{Berkeley Institute for Data Science, University of California, Berkeley, CA 94720, USA}

\begin{abstract}
Most of the fundamental, emergent, and phenomenological parameters of particle and nuclear physics are determined through parametric template fits.  Simulations are used to populate histograms which are then matched to data.  This approach is inherently lossy, since histograms are binned and low-dimensional. Deep learning has enabled unbinned and high-dimensional parameter estimation through neural likelihiood(-ratio) estimation.  We compare two approaches for neural simulation-based inference (NSBI): one based on discriminative learning (classification) and one based on generative modeling.  These two approaches are directly evaluated on the same datasets, with a similar level of hyperparameter optimization in both cases.  In addition to a Gaussian dataset, we study NSBI using a Higgs boson dataset from the FAIR Universe Challenge.  We find that both the direct likelihood and likelihood ratio estimation are able to effectively extract parameters with reasonable uncertainties.  For the numerical examples and within the set of hyperparameters studied, we found that the likelihood ratio method is more accurate and/or precise.  Both methods have a significant spread from the network training and would require ensembling or other mitigation strategies in practice.
\end{abstract}

\maketitle
\flushbottom

\section{Introduction}
\label{sec:intro}

At the level of reconstructed particle properties, collider physics events are high dimensional. There can be many outgoing particles, each with a four-vector and other properties like electric charge.  Traditionally, these complex data are analyzed by first compressing the many-dimensional phase space into a small number of high-level features and then discretized.  Histograms filled with simulated data are then compared with experimental data.  The parameters of the simulation that produce the best match are declared the fitted values.  While highly successful, this approach under-utilizes the available information.

Modern machine learning (ML) provides an alternative approach that can process the full phase space holistically.  Instead of using high-level features and histograms, neural simulation-based inference (NSBI) allows for users to perform a likelihood-based analysis without an explicit form of the likelihood~\cite{Cranmer_2020}.  There are many methods for performing NSBI.  Two well-studied approaches are discriminative (classifier-based) methods and generative methods.  Discriminative methods use machine learning tools to approximate likelihood ratios $p(x|\theta)/p(x|\theta_0)$ for features $x$ and parameters $\theta$.  The parameter $\theta_0$ is a constant, so maximizing the ratio is equivalent to maximizing the numerator.  In contrast, generative models use machine learning techniques to directly approximate $p(x|\theta)$.  Many approaches actually approximate $p(\theta|x)$, but we will take a frequentist approach and focus on likelihoods.  There is an analogous Bayesian interpretation in terms of posteriors.  We note that while NSBI can be used for a number of applications (e.g. unfolding~\cite{Arratia:2021otl,Butter:2022rso2,Huetsch:2024quz}), we focus on the well-studied case of parameter estimation.

NSBI tools for parameter estimation (henceforth, just NSBI) have been extensively studied phenomenologically and there are a growing number of experimental measurements using these methods in collider physics~\cite{ATLAS:2024jry,ATLAS:2024ynn,CMS:2024ksn} and related areas~\cite{Hermans:2020skz,Dax:2022pxd}.  Each study focuses on discriminative or generative approaches to NSBI.  We are not aware of a direct comparison between these two approaches\footnote{While this manuscript was being finalized, Ref.~\cite{chatterjee2025exploringbsmparameterspace} appeared on the arXiv.  It similarly compares these approaches, but uses different comparison metrics and features a different approach for integrating the signal-rate into the overall likelihood calculations. }.

The goal of this paper is to directly compare discriminative and generative NSBI techniques on a common dataset.  To benchmark this comparison, we consider the problem of Higgs boson characterization at the LHC.  This is the focus of the community-wide FAIR Universe HiggsML Uncertainty Challenge~\cite{Bhimji:2024bcd}, which builds on the successful HiggsML Challenge~\cite{pmlr-v42-cowa14} by integrating uncertainty quantification.  While we suspect that the optimal NSBI method is application-specific, the Higgs boson characterization case is of inherent interest and it is similar to a number of related problems so lessons learned may be useful more generally.

%We should be upfront with our limiations, e.g. need an exact background model within uncertainty.

This paper is organized as follows.  Section~\ref{sec:methods} introduces NSBI for parameter estimation.  The datasets we use for comparisons are described in Sec.~\ref{sec:data} and the metrics used for evaluating each method are detailed in Sec.~\ref{sec:eval}.  Numerical results are presented in Sec.~\ref{sec:results} and the paper ends with conclusions and outlook in Sec.~\ref{sec:conclusions}.

\section{Methods}
\label{sec:methods}

In this work, we will compare two complementary approaches to NSBI, one that uses a classifier machine learning (ML) model and one that leverages generative ML models.  In the context of frequentist inference, the goal of any NSBI method is to implicitly of explicitly approximate the likelihood function 
\begin{equation}
    p(x|\mu, z)\;,
\end{equation}
where $x$ is the set of measured data, $\mu$ is the parameter(s) of scientific interest, and $z$ is a set of nuisance parameter(s) that affect the likelihood function, but are themselves not of interest.  The goal is to maximize the likelihood over $\mu$ and $z$, marginalize over $z$, and report confidence intervals around $\mu$ given by the shape of the likelihood around the maximum.

We will focus on a common setting for particle physics: the data are a mixture model of a signal process and a background process.  The parameter of interest is the fraction of signal in data.  Systematic uncertainties are controlled by nuisance parameters that affect both the signal and background probability densities.

\subsection{Direct Likelihood Estimation}

We start with directly approximating the probability density $p(x|\mu,z)$.  One machine learning tool well-suited for this task is the normalizing flow~\cite{papamakarios2021normalizingflowsprobabilisticmodeling}.  Normalizing flows are invertible functions that implement the change of variables formula.  Typically, a normalizing flow starts with a standard normal random variable and then maps it to the data space.  The map is composed of a series of invertible transformations with a computationally tractable Jacobian so that one can compute the probability density of the composition.  The normalizing flow is optimized by maximizing the probability density in the data space.

When a normalizing flow is trained to learn the parameters given the data, the approach is called \textit{neural posterior estimation}.  In frequentist analysis, we instead learn the probability density of the data given the parameters and then consider the result as a function of the parameters for fixed data.  This can be achieved with neural posterior estimation using a uniform prior or by training a conditional normalizing flow.  We use the latter approach.

It is challenging to learn the full likelihood directly because each event has so little information about the signal strength $\mu$.  Instead, we use the mixture model nature of the problem to train two normalizing flows - one for the signal and one for the background:

\begin{equation}
\label{eq:nf_mixture}
    p(x|\mu, z) = \frac{\mu}{\mu+1}p_\text{sig.}(x|z) + \frac{1}{\mu+1}p_\text{back}(x|z)\,,
\end{equation}
where the signal and background probability densities are trained to be conditional on the nuisance parameter $z$. 

%Normalizing flow models are directly able to learn the likelihood of points in a given dataset, and as a result have seen use for Neural Posterior Estimation (NPE) in field such as astrophysics[Todo:add cite]. However, using flows for NPE in particle physics come with additional challenges, as information such as the signal rate $\mu$ cannot be determined from individual events, and are instead only determinable from a set of multiple events. Therefore, it is not possible to use a flow conditioned on $\mu$ to determine $p(x|\mu)$. Instead, we train two separate normalizing flows, one that approximates $p(x|sig,z)$ and one that approximates $p(x|bkg,z)$, where $x$ is the event vector in feature space. These are then mixed with a signal ratio $\mu$ to obtain 
%\begin{equation}
%    p(x|\mu, z) = \frac{\mu}{\mu+1}p(x|sig, z) + \frac{1}{\mu+1}p(x|bkg, z)
%\end{equation}

\subsection{Likelihood Ratio Estimation}

An alternative approach to approximating $p(x|\mu,z)$ is to estimate the ratio $p(x|\mu,z)/p(x|\mu_0,z_0)$ for fixed values $\mu_0,z_0$.  The reason for considering the ratio is that this converts the problem from density estimation to classification.  Training a classifier to distinguish two samples is usually set up to learn the probability of the first sample.  If the two samples are indexed with $(\mu_1,z_1)$ and $(\mu_0,z_0)$, this means that we can extract the likelihood ratio from the classifier $f(x)\approx \Pr(\mu_0,z_0|x)$:
\begin{align}\nonumber
    \frac{p(x|\mu_1,z_1)}{p(x|\mu_0,z_0)}&=\frac{\Pr(\mu_1,z_1|x)\,p(x)/\Pr(\mu_1,z_1)}{\Pr(\mu_0,z_0|x)\,p(x)/\Pr(\mu_0,z_0)}\\\label{eq:lltrick}
    &\approx\frac{f(x)}{1-f(x)}\,\frac{\Pr(\mu_0,z_0)}{\Pr(\mu_1,z_1)}\,,
\end{align}
where $p(\cdot)$ denotes a probability density and $\Pr(\cdot)$ represents a probability mass.
When there are equal numbers of the two samples in training $f$, then $\Pr(\mu_1,z_1)=\Pr(\mu_0,z_0)$ and the likelihood ratio is approximated by $f(x)/(1-f(x))$, a fact that has been well-known in particle physics for many years~\cite{cranmer}.  Similar to the previous section, we can use the mixture model nature of the problem to break out the $\mu$ dependence:

\begin{align}
\label{eq:llr_mm}
    \frac{p(x|\mu, z)}{p(x|\mu_0, z_0)} = \frac{\mu}{\mu+1}\frac{p_\text{sig.}(x|z)}{p(x|\mu_0, z_0)} + \frac{1}{\mu+1}\frac{p_\text{back.}(x|z)}{p(x|\mu_0, z_0)} \; .
\end{align}
The continuous likelihood ratios are approximated using parameterized classifiers~\cite{cranmer,Baldi:2016fzo}.  Instead of training a classifier with input $x$ to distinguish samples drawn from two discrete set of parameters, a classifier is trained on $(x,z)$ to distinguish one sample where each event has a $z$ drawn from a distribution $p(z)$ while the other sample is generated with a fixed $z=z_0$, but then the network is presented $(x,z)$ where $z$ is randomly drawn also from $p(z)$.  This gives
\begin{align}
    \frac{f(x,z)}{1-f(x,z)}\approx\frac{p(x,z)}{p(x|z_0)p(z)}=\frac{p(x|z)}{p(x|z_0)}\,.
\end{align}
which is what we need for the two likelihood ratio terms in Eq.~\ref{eq:llr_mm}.

\subsection{Implementation}

All normalizing flows and classifiers are parameterized as neural networks and implemented in \textsc{PyTorch}~\cite{NEURIPS2019_9015}. 
The signal and background classifier models have an identical architecture, consisting of a 3-node input layer, 3 hidden layers with 120 nodes, and a final 1-node output layer. Each layer has a standard \textsc{ReLU} activation function, except for the output layer, which uses a Sigmoid activation.  
Both flow models are conditional flows with 2 input and output dimensions. 
The conditional input consists of one dimension, which gets expanded into 32 dimensions via a fully connected embedding layer. 
Each model consists of 4 autoregressive masked piecewise rational quadratic spline~\cite{durkan2019neural} blocks.
For each block, the spline parameters are learned by a fully connected network with two 64-node layers and \textsc{ReLU} activation functions.
The network architecture hyperparameters for both model classes were selected based on reasonable baselines from the literature and were not extensively optimized. The training hyperparameters were chosen similarly, with the exception of the learning rate, which was found to have a noticeable impact on the model performance, and, as a result, was optimized using a logarithmic scan.

\subsection{Inference}

As all events are independent, the likelihood for a full dataset is the product of likelihoods across events~\cite{Nachman:2021yvi}.  We improve the numerical stability by taking the logarithm of the likelihood or likelihood ratio so that the product across events is instead a sum over events.  

For the inference step, we freeze the neural network weights and only vary $\mu$ and $z$.  We could use the differentiability of neural networks to directly optimize for $\mu,z$ with the network weights fixed.  For this paper, we do a simple grid search, first profiling over $z$, and then performing a fourth-order polynomial to interpolate between grid points. This works well in the low-dimensional examples that follow, but the automatic differentiation approach is likely required when more parameters are included.  Confidence intervals are constructed by identifying where the log likelihood decreases by $0.5$ from the maximum.

\section{Data Sets}
\label{sec:data}

We compare Direct Likelihood Estimation (DLE) and Likelihood Ratio Estimation (LRE) using two datasets - one based on Gaussians and one built from a collider physics example.

%We compare the performance of the generative-based and classifier-based NPE approach on two datasets. 

\subsection{Gaussian Example}
\label{data:gauss}

\begin{figure}[h]
    \centering
    \includegraphics[width=0.49\linewidth]{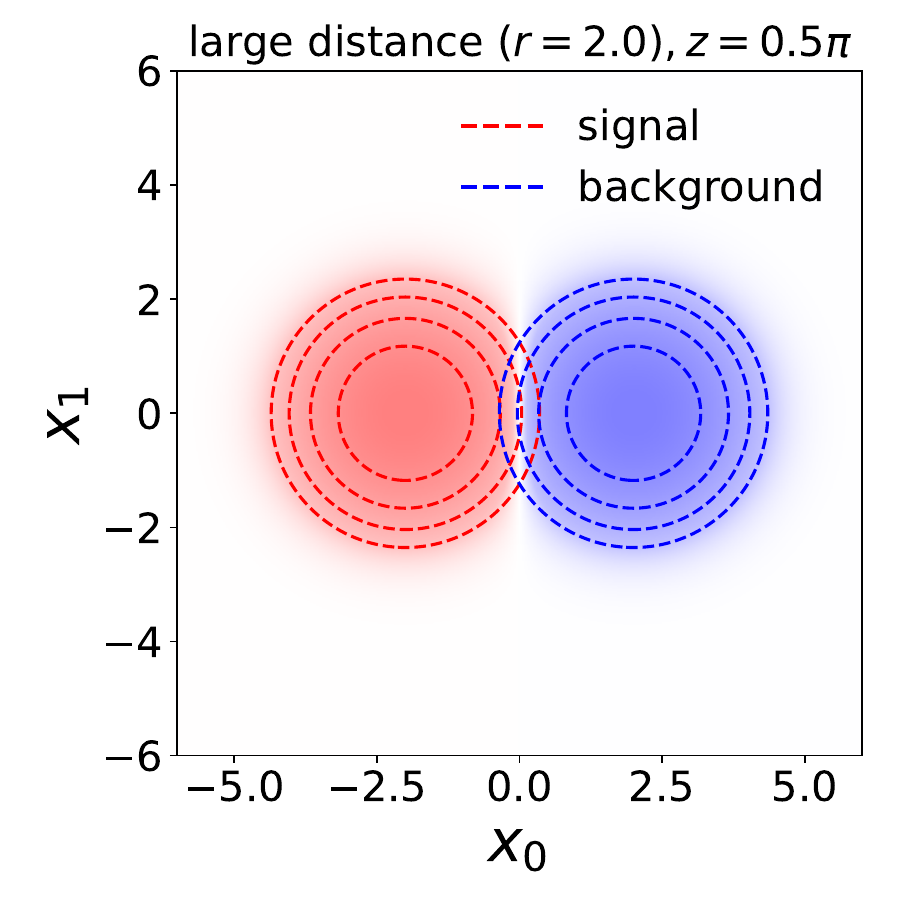}
    \includegraphics[width=0.49\linewidth]{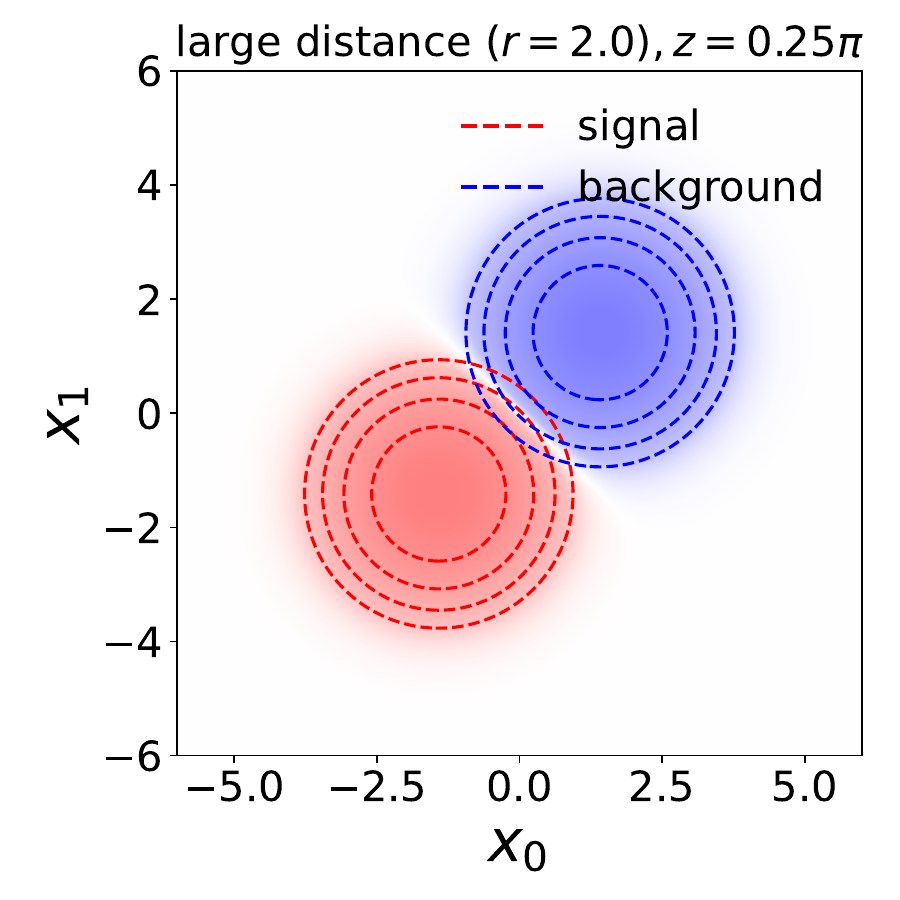}
    \includegraphics[width=0.49\linewidth]{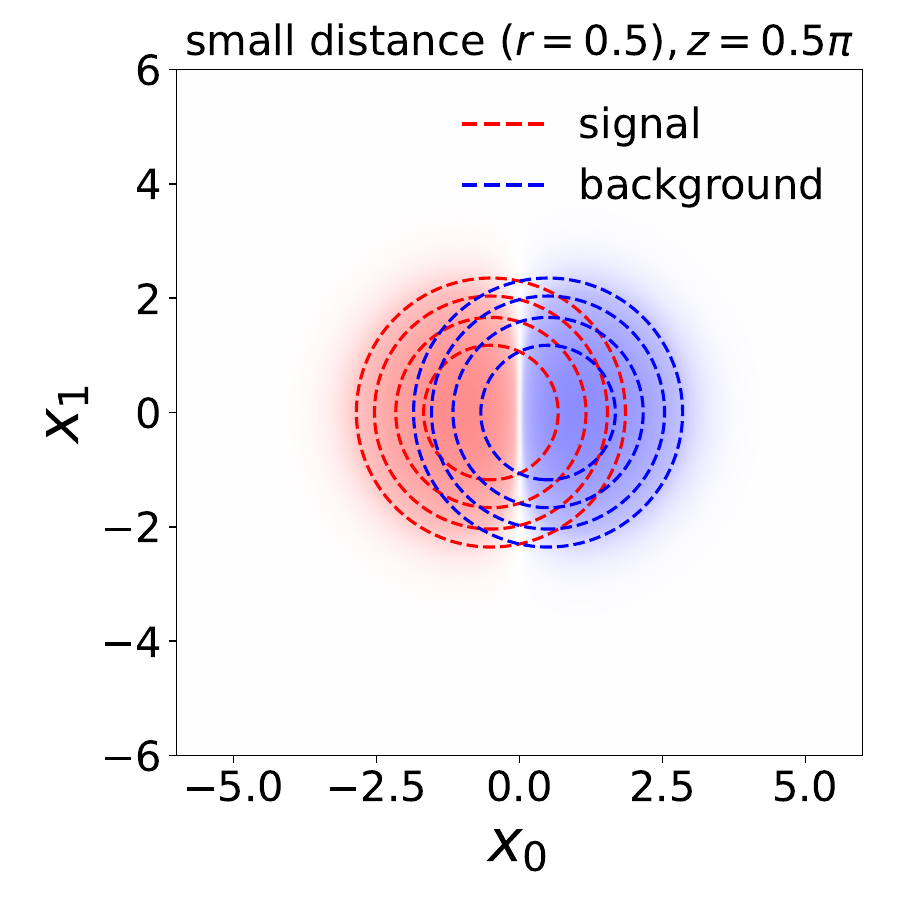}
    \includegraphics[width=0.49\linewidth]{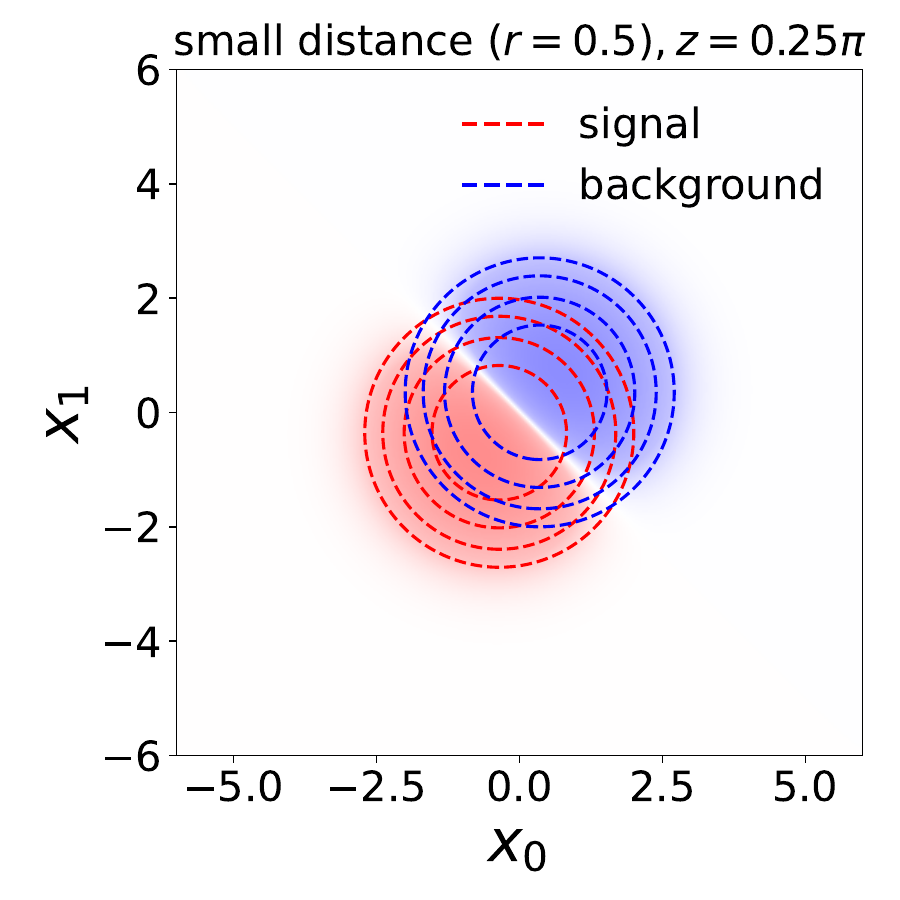}
    \caption{Vizualizsation of the Gaussain data for rotation values $z=0.5$ (left-hand side) and $z=0.25 \pi$ (right-hand side), as well as for the large distance case (upper panels) and the small distance case (lower panels). }
    \label{fig:2d_features}
\end{figure}

Our preliminary tests are performed on a synthetic dataset consisting of two, two-dimensional Gaussians. One of the Gaussians is designated as `signal', while the other is designated as `background'. The distance between the Gaussian mean and the origin is given by the factor $r$. As a result, the separation between the two means is $2r$. We distinguish between two cases: 
\begin{itemize}
    \item The long-distance case with $r=2.0$
    \item The short-distance case with $r=0.5$
\end{itemize}
Since the difficulty of classifying a given event as either signal or background is directly related to the overlap between the Gaussians, and therefore $r$, this enables us to test the model performance on both simple and challenging classification tasks. 
We additionally introduce a systematic nuisance parameter $z$, which corresponds to the rotation of the means of the Gaussians around the origin~\cite{Ghosh:2021roe}. 
The full definition of the dataset is given by:
\begin{align}
    X_\text{back.} &\sim (\mathcal{N}(\cos(z)r, 1),\mathcal{N}(\sin(z)r, 1)) \\
    X_\text{sig.} &\sim (\mathcal{N}(-\cos(z)r, 1),\mathcal{N}(-\sin(z)r, 1)) 
\end{align}
where $\sim\mathcal{N}(\alpha,\beta)$ represents a random variable that is normally distributed with mean $\alpha$ and variance $\beta$.
The Gaussian data are illustrated in Fig~\ref{fig:2d_features} with four combinations of $r$ and $z$ values. 

A training set for the Gaussian data consists of one million events, evenly split between signal and background. For each point, a nuisance value was uniformly sampled from $z \in [0, 0.5\pi]$. To facilitate the training of the classifier models, an additional training set of the same size was generated, with a fixed nuisance value of $z_{0}=0.25\pi$, to act as a reference for the likelihood ratio calculation. Additional validation sets were directly generated during the model training. A test set for evaluation consists of 9000 total data points with fixed $\mu$ and $z$ values. 

The main reason to study the Gaussian data is that the true likelihood function is known, so we can compare the learned likelihood (ratio) with the true values.  For the physics case presented in the next section, the true likelihood (ratio) is not known.

\subsection{Higgs boson Example}

The Higgs boson was the last particle of the Standard Model (SM) to be discovered~\cite{ATLAS:2012yve,CMS:2012qbp}.  While its properties are highly constrained within the context of the SM, the Higgs boson also plays a central role in many theories of physics beyond the SM (BSM).  In such BSM theories, the Higgs properties vary from the SM ones and so searching for deviations is a central task in collider physics.  Many papers have been written on the use of machine learning for studying Higgs boson properties and a community data challenge on this subject attracted attention from researchers outside of particle physics~\cite{pmlr-v42-cowa14}.  Now that the existence of the Higgs boson is well-established, there is a need to focus on precision and so the FAIR Universe HiggsML Uncertainty project was started to extend the earlier data challenge by bringing in uncertainty quantification~\cite{Bhimji:2024bcd}.  We use a version of the FAIR Universe HiggsML Uncertainty dataset in this paper, which is briefly described below.

The simulated data consist of semi-leptonic $H \to \tau \tau$ signal events, and a similarly semi-leptonic $Z \to \tau \tau$ background. The data were simulated using \textsc{Pythia 8.2}~\cite{Sjostrand:2014zea}, and \textsc{Delphes 3.5.0}~\cite{deFavereau:2013fsa} for the detector simulation. 

As both signal and background events feature a hadronically decaying $\tau$ in their final state, this dataset is highly sensitive to the energy calibration scale used for the $\tau$ jets (tau energy scale or TES). This makes the TES a well-suited nuisance parameter to test the ability of the machine learning models to handle a real systematic effect. The effects of the TES are not included in the initial simulation but are modeled ad-hoc during the model training and evaluation. 

\begin{figure}[h]
    \centering
    \includegraphics[width=0.49\linewidth]{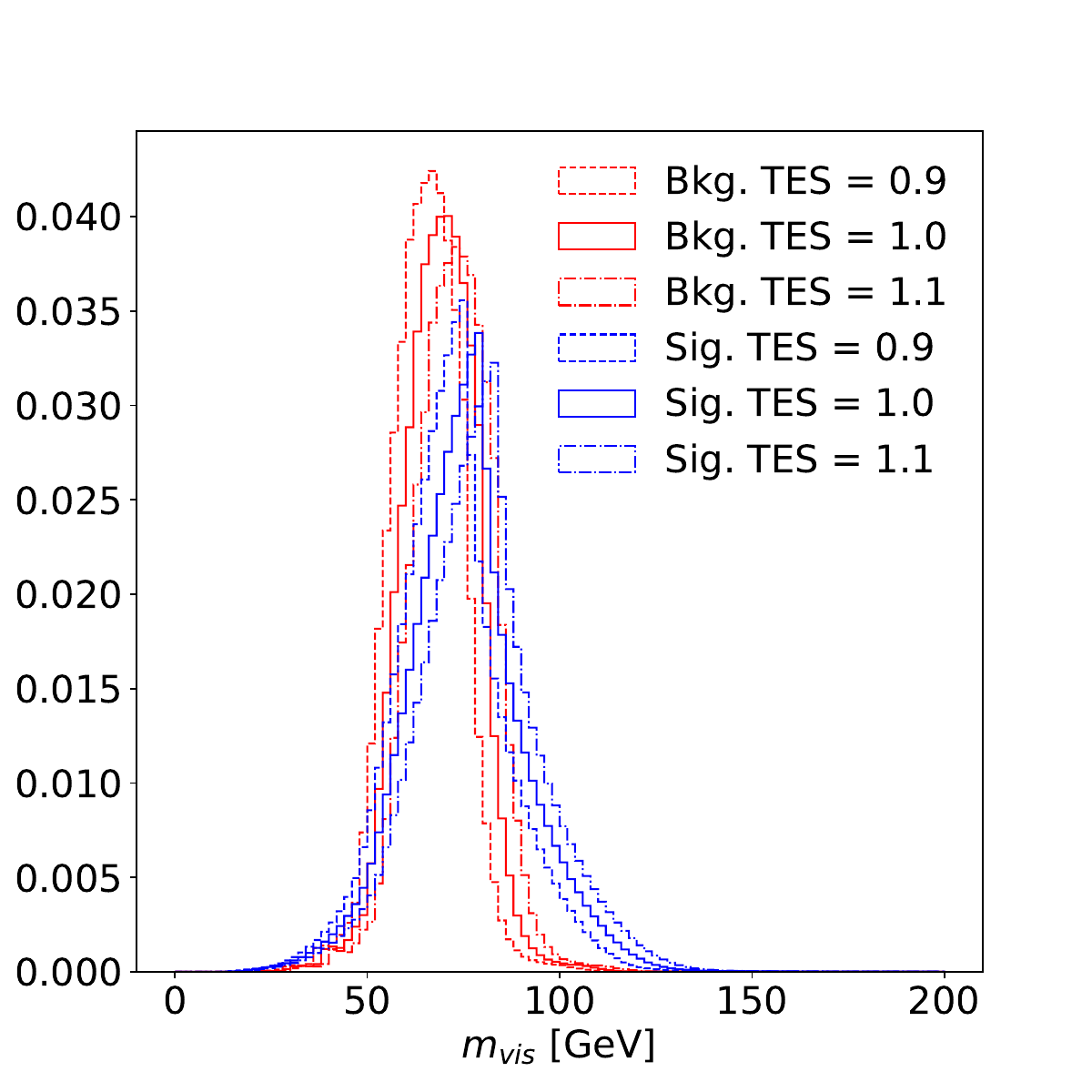}
    \includegraphics[width=0.49\linewidth]{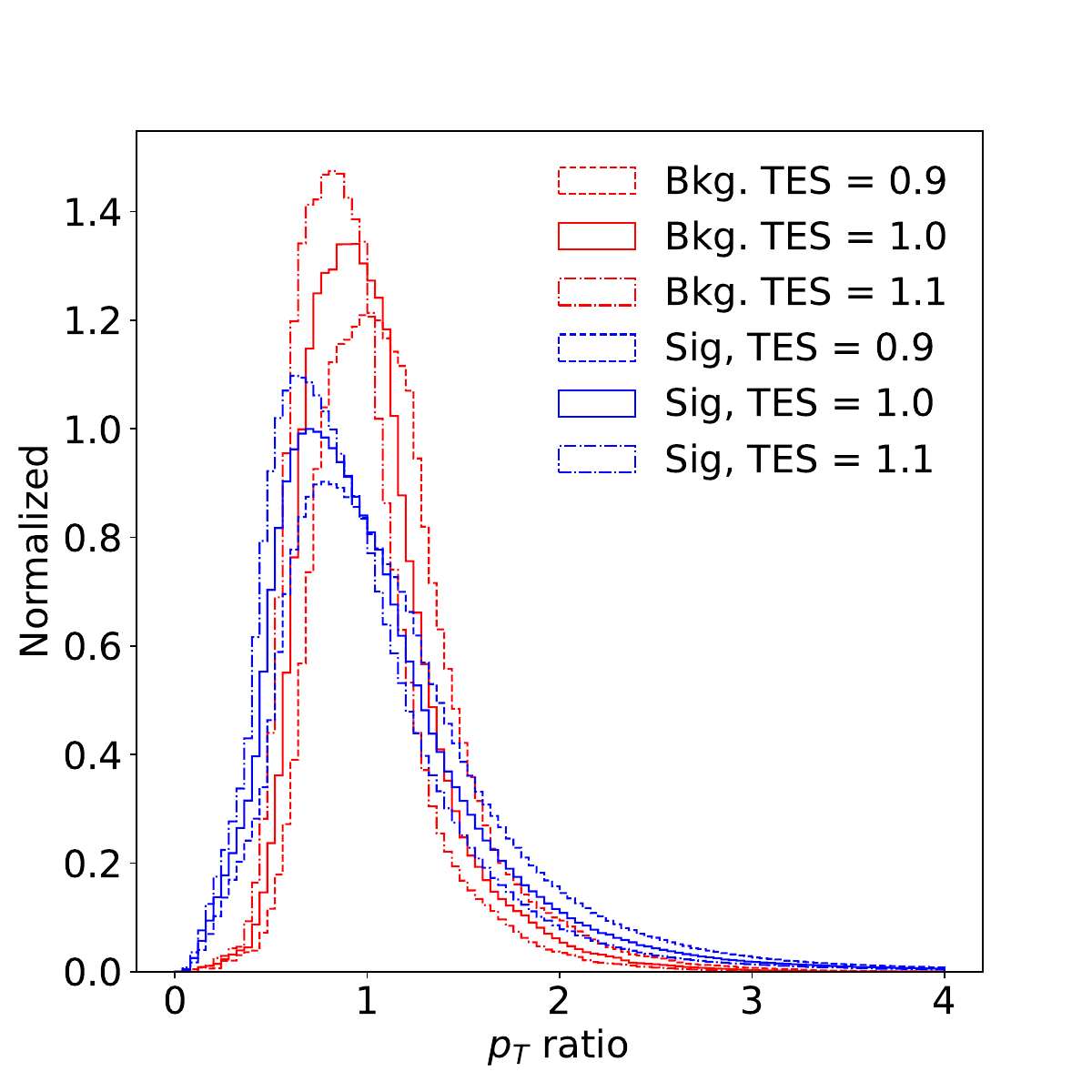}
    \caption{Visualization of the features used for the Higgs data. The left-hand plot shows the invariant mass of the hadronic $\tau$ and non-$\tau$ lepton, the right-hand plot shows the $p_{T}$ ratio of the hadronic $\tau$ and non-$\tau$ lepton. The solid lines show the features for the nominal TES value, while the dashed and dash-dotted lines show them for smaller and larger TES values, respectively. }
    \label{fig:higgs_features}
\end{figure}

To reduce the computational cost associated with the many pseudoexperiments we will perform later, we select two observables to use in the model training. Specifically, we chose the invariant mass of the hadronic tau and the non-$\tau$ lepton ($m_\text{vis}$), and the ratio of transverse momenta ($p_{T}$) between the hadronic tau and the non-$\tau$ lepton. Both of these observables are different for signal and background and are highly sensitive to the TES, making them ideal candidates to test the systematic effects. Histograms of the two features are illustrated in Fig~\ref{fig:higgs_features}. 

Unlike the Gaussian data, the Higgs data requires dedicated simulation software to generate data points. Therefore, we are limited to the existing simulation data and cannot simply generate more points on demand. As a result, we allocate the available data in the following way:
\begin{itemize}
    \item Train set: 7 million events total, divided into 2.9 million background and 4.1 million signal events\footnote{The class imbalance that leads to the factor in Eq.~\ref{eq:lltrick} is a constant and does not affect the log likelihood optimization.},
    \item Validation set: 400,000 events total, split evenly between signal and background,
    \item Test sets: 6.02 million events total, with 6 million background and 22,500 signal events.
\end{itemize}
The class split emulates the small signal-to-background ratio found in realistic physics data as all events have unit weight, while the more even split in the training set is in line with the common practice to over-sample the signal case in particle physics simulation.

\section{Evaluation}
\label{sec:eval}

\subsection{Uncertainty Quantification Performance}

Unlike standard classification tasks, we are not interested in the raw performance of a given model.  Instead, we want confidence intervals to be as small as possible while also correctly representing the spread expected if we re-ran the experiment many times.
%
%One unique challenge faced by uncertainty-aware machine learning methods is that the quality evaluation of a model goes beyond pure classification performance. Most notably, one needs to ensure the confidence interval predicted by the method correctly represents the uncertainty that would be observed by re-running the underlying measurement multiple times. 
%
In order to emulate multiple experiments, each time we would evaluate a model on a test set, we instead evaluate the model multiple times on bootstrapped re-samplings of the test set~\cite{10.1214/aos/1176344552}. For $N$ bootstrap samplings, this results in $N$ confidence intervals. Since each confidence interval represents $1\sigma$, we expect that about $68.2\%$ of the intervals will contain the true value. Therefore, we calculate the fraction of bootstrap evaluations for which the known $\mu$ and $z$ values of the test set lie within the predicted interval. We refer to this percentage as the \textit{coverage}.  Since we also want precision, a second metric is the average size of the $1\sigma$ intervals, with smaller sizes representing better performance.

%This coverage metric forms a fundamental requirement for any uncertainty-aware machine learning model, as large deviations away from $68.2\%$ indicate that the uncertainties estimated by the model are not reliable. However, the coverage metric is agnostic to the size of the predicted $1\sigma$ interval, so long as the spread of predicted intervals agrees with interval size. This means that a method can have perfect coverage, but still produce a very large uncertainty range. In a real particle physics application, however, there is a strong incentive to minimize the size of the uncertainty interval, as this corresponds to more precise measurements. Therefore we introduce a secondary performance metric consisting of the full with of the $1\sigma$ interval, where a lower width value corresponds to a better model performance. 

For the Gaussian dataset, we introduce another metric because the true likelihood is known.  The overlap metric $o$ is defined as
\begin{align}
    o = 1 - \frac{1}{2} \sum_{\theta\in\Theta} |\hat{p}_\text{norm}(\{x\}|\theta) - p_\text{norm}(\{x\}|\theta)| \; ,
\end{align}
where
$\Theta$ is a grid over either $\mu$, $z$, or $(\mu,z)$ (if only one, the other is marginalized over), $\hat{p}$ is the estimated likelihood (ratio), $p$ is the true likelihood, $p(\{x\}|\theta)=e^{-\text{LL}}$ for log likelihood (ratio) LL over all events up to a constant independent of $\theta$, and $p_\text{norm}(\{x\}|\theta)=p(\{x\}|\theta)/\sum_{\theta'\in\Theta} p(\{x\}|\theta')$.  The normalization allows the likelihood ratio methods to be directly comparable to the likelihood methods.

This gives us a metric between 0 and 1, which allows us to directly gauge how well the model-predicted likelihood agrees with the true likelihood of the Gaussian data set.  

%$NLL(x)_{i_{\mu}, i_{z}}$, where $N$ and $M$ are the extents of the grid, and $i_{mu} \in [1,N]$, $i_{z} \in [1,M]$

\subsection{Evaluation Process}

Since the proposed coverage metric is only applicable in aggregate, we require the method to be evaluated on many evaluation sets to be able to quantify the spread of predicted intervals. To this end, we make use of data splitting and bootstrapping.  While the natural signal-to-background ratio is much smaller, we consider $\mu\sim\mathcal{O}(10\%)$ in order to facilitate faster training to more readily perform the full battery of coverage tests. The 22,500 signal events in the test set are split into 5 disjoint sets.  For each set, we have a total of $9000$ events, which is approximately the maximum allowable for the highest signal fraction we scan (30\%) and fixing the number of background events.  From each set of $9000$, we re-sample with replacement $1,000$ points (explained in Sec.~\ref{sec:k})). For each $9,000$ point set, we draw a total of 100 bootstrapped datasets. A given model is then evaluated on these bootstrapped sets, and the resulting confidence intervals are used to calculate the coverage metric.  The computational bottleneck is from the optimization and confidence interval finding for a given dataset due to the fine scan in $\mu$ and $z$.  It may be possible to accelerate this with gradient descent and approximate confidence intervals based on the Fisher matrix.  The number of bootstrapped datasets is not limiting because we evaluate the network on all data points before (sub)sampling.  Due to the averaging across disjoint sets and across parameter values (described below), we did not find a significant improvement in precision by creating more pseudodatasets.   

Additionally, we require multiple $9,000$ point starting evaluation sets, generated with different values for $\mu$ and $z$ in order to ensure the robustness of a given method under changes to these values. Therefore, we define a grid in $\mu$, $z$ with $\mu \in [0.1, 0.2, 0.3]$ for both datasets and $z \in [0.15\pi, 0.25\pi, 0.35\pi]$ or $z \in [0.93, 1.00, 1.07]$ for the Gaussian and Higgs data, respectively. For each of these 9 grid points, we create a test set, either by generating a new point in the Gaussian case, or by selecting an appropriate number of signal and background events from the set-aside test set in the Higgs case. 

The entire bootstrapping process is repeated on each of the five disjoint sets, resulting in 45 evaluation sets.  For each bootstrapped dataset, we train $N$ models ($N=30$ for the Gaussian and $N=40$ For the Higgs).  In summary: for each of the 45 evaluation sets and each of the $N$ models, we calculate the relevant metrics over the 100 bootrapped datasets and average for each of the $N$ models.  In practice, one would could make the method more precise by ensembling the $N$ models.  We study their spread as a way to illustrate the method robustness to random network initalizations.  Even though this spread can be reduced through averaging, it does add to the computational complexity of the setup.  It may be also be possible in the future to mitigate this spread through more extensive hyperparameter optimization, including using more robust optimizers~\cite{DeLuca:2025ruv}.

\subsection{Bootstrapping Evaluation}
\label{sec:k}

For a given set of parameters, the previous section described how confidence intervals are obtained using bootstrapped datasets.  We use the simplest estimate of the 68\% confidence interval from these estimates - $\hat{\theta}_{(84)}-\hat{\theta}_{(16)}$, where $\hat{\theta}$ is the predicted value of the parameter $\theta$, and the subscript with parentheses denotes the order statistics from the 100 bootstrapped pseudodatasets.  There are other, more accurate estimates~\cite{10.1214/ss/1177013815} that it would be interesting to study in future work. All of these estimates have the property that they converge to the true confidence interval as the number of original samples grows to infinity.  In the finite dataset limit, the confidence interval estimates can be biased.  We found that one way to reduce the bias is to sample $M$ events from $kM$ events ($k>1)$ with replacement instead of sampling $M$ from $M$.  This section briefly explores how big to make $k$.

%While developing the evaluation chain outlined above, several non-trivial interactions between the bootstrapping procedure were encountered. To further quantify these effects, we ran the full evaluation chain while varying both the size of the bootstrapped subsets between $1,000$ and $5,000$, as well as the ratio $k$ between the size of the evaluation set and the size of the subset.  For example, an evaluation set size of $25,000$ and a subset size of $5,000$ would correspond to $k= \frac{25,000}{5,000} = 5$. 

\begin{figure}[hbt]
    \centering
    \includegraphics[width=0.95\linewidth]{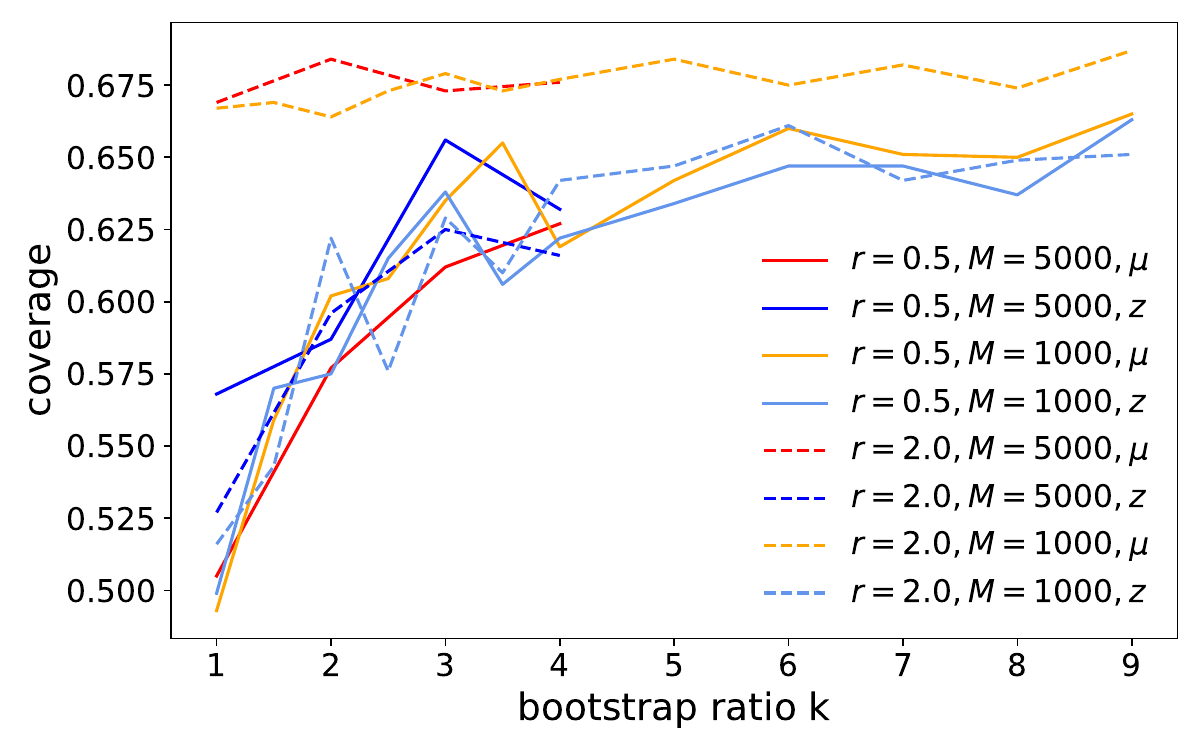}
    \caption{Coverages vales as a function of the ratio $k$ between the size of an evaluation set and the size of the bootstrapped subset ($N$).  Inference is performed on either $\mu$ or $z$ as indicated in the legend. Results are shown for both the small-distance ($r=0.5$) and large-distance ($r=2$) Gaussian datasets.}
    \label{fig:boostrapping_test}
\end{figure}

For this study, we use the Gaussian dataset, using the true Gaussian likelihoods for inference, to remove potential biases introduced from the ML modeling.
%
%All tests were performed in the Gaussian toy set, using the know, true Gaussian likelihood, to remove effects introduced by imperfect ML modeling. 
The results are shown in Fig.~\ref{fig:boostrapping_test}.  The coverage from 100 bootstrapped datasets is shown as a function of $k$, for $M\in\{1000,5000\}$, $\theta\in\{\mu,z\}$, and for the small ($r=0.5$) and large ($r=2$) distance configurations from Sec.~\ref{data:gauss}.
% 
%Different line colors correspond to different subset sizes, and to whether the coverage metric was evaluated for $\mu$ or $z$. Solid and dashed line styles correspond to the small and large distance Gaussian examples, respectively. 
%
There does not seem to be a noticeable difference in the behavior between subsets with size $1,000$ and subsets with size $5,000$, as the corresponding pairs of lines overlap within their fluctuations. 
As a result, cases with subsets-size $5,000$ and large values of $k > 4$ were left out, as testing them would have required significant computational cost without likely contributing additional insight. 
Only the large distance Gaussian case for inferring $\mu$ reaches the expected coverage of about $68\%$ for all values of $k$. 
Meanwhile, all coverages for the small distance case, as well as all results involving $z$, require a comparatively large $k > 5$ to approach the expected coverage. 

For this reason, we settled on using a subset size of $1,000$ and $k=9$ in our evaluation chain.  A residual bias may still exist, which may be responsible for small variations about from 68\% coverage in the final results.
%It should also be noted, that small variations in the coverage away from $68\%$ in the final results for the models that were investigated can potentially be attributed to this effect. 

\section{Results}
\label{sec:results}

\subsection{Gaussian Example}

\begin{figure*}[ht]
    \centering
    \includegraphics[width=0.32\textwidth]{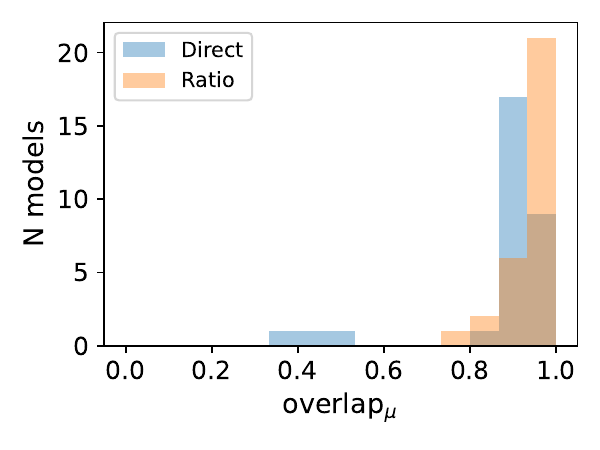}
    \includegraphics[width=0.32\textwidth]{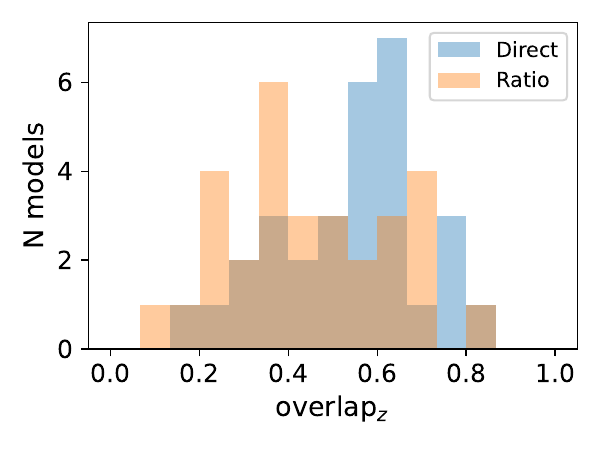}
    \includegraphics[width=0.32\textwidth]{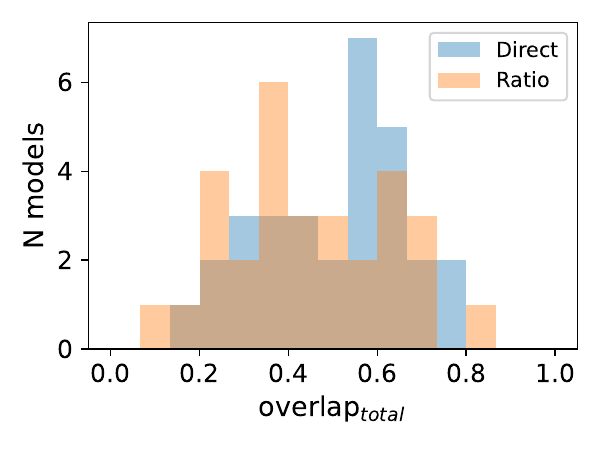}
    \includegraphics[width=0.34\textwidth]{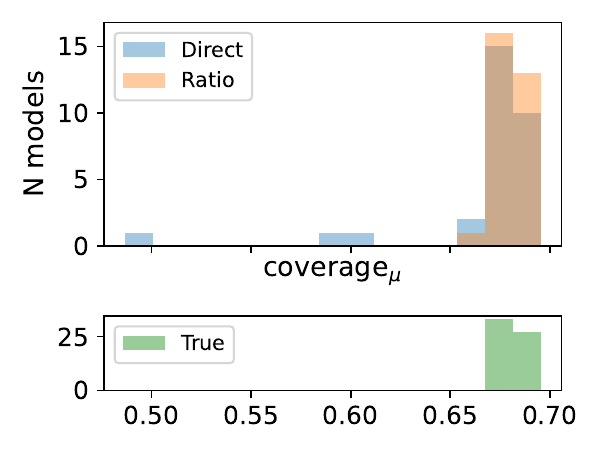}
    \includegraphics[width=0.34\textwidth]{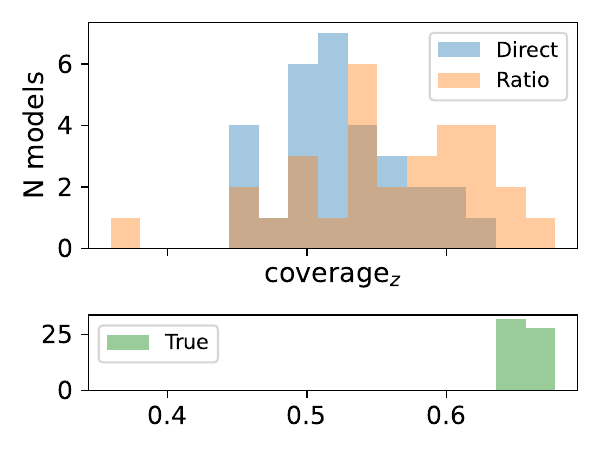}
    \includegraphics[width=0.34\textwidth]{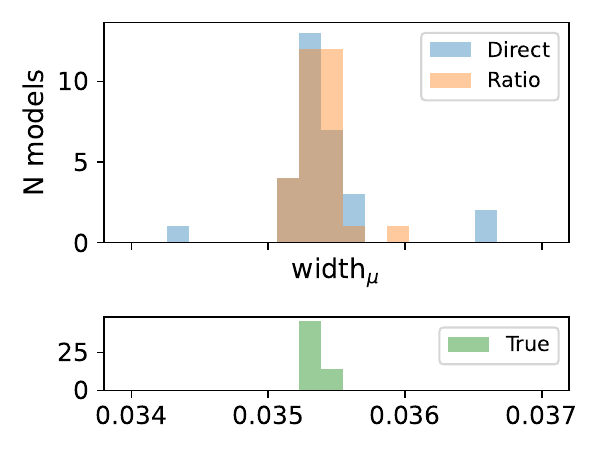}
    \includegraphics[width=0.34\textwidth]{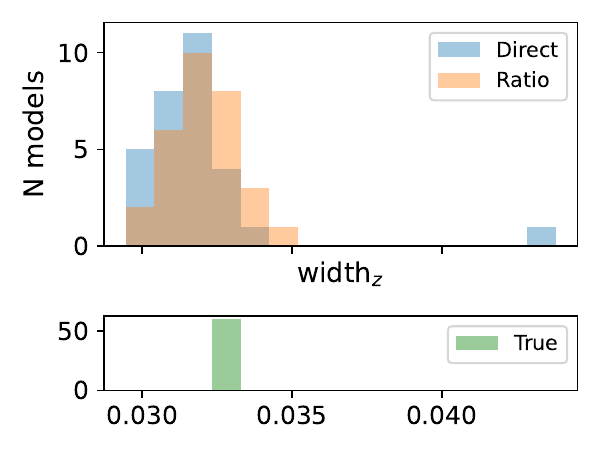}
    \caption{Gaussian long distance case results for the \flow (learning rate = $3\times 10^{-6}$) and \classi model (learning rate = $3\times 10^{-5}$). Histograms show the performance of 30 independently trained models of both the \classi model and the \flow model. For the coverage and mean width results, the plots are split into an upper section for the model results and a lower section for the ground truth model, indicating approximately what the histograms should look like if all 30 models were the perfect model.}
    \label{fig:hist_large_dist}
\end{figure*}

For both the small and large distance Gaussian data sets, $N=30$ \flow and \classi models were trained and evaluated as described in Sec.~\ref{sec:eval}. 
The results of this for the large distance case are shown in Fig.~\ref{fig:hist_large_dist}. 
The three top-row panels correspond to the overlap metrics between the model-predicted likelihoods and the true Gaussian likelihood. 
The overlap in the $\mu$ likelihood is shown in the right-hand panel, the overlap in $z$ is shown in the center panel, and the left-hand panel shows the combined overlap in $\mu$ and $z$. 
The overlap for $\mu$ is narrowly peaked around $1.0$ for both the \flow and \classi models, with only a small number of outliers for both models, indicating that both models perform adequately for determining $\mu$. 
The $z$ overlap is clustered around $0.5$ and notably broader. 
This indicates that extracting the precise profile likelihood in $z$ is a more challenging task compared to $\mu$, which is in line with intuitive expectations for the large distance case. 
As a result, the combined overlap is largely dominated by the mismatches in $z$, and is therefore nearly identical to the $z$ overlap. 
The center and lower rows show the coverage and width metrics, respectively, split into the results for $\mu$ in the left-hand column and the results for $z$ in the right-hand column. 
Each plot shows the distribution of metric values for the $N$ models in the top section and the equivalent results obtained using the true Gaussian likelihood in the bottom section. 
This allows us to determine if any deviation in the coverage can be linked to the bootstrapping effects investigated in Sec.~\ref{sec:eval} or if it is caused directly by the model performance.
The results of the coverage are in line with what was observed in the overlap; for $\mu$, we see excellent coverage, in line with what is obtained using the true likelihood, while there are notable mismatches in $z$.
Finally, the predicted interval widths of the models are in line with the true likelihood, indicating that for this dataset, both models perform very well in determining $\mu$. 
It should be noted that the widths obtained by the models in $z$ do not deviate significantly from the widths obtained from the truth. 
This indicates that the imperfect coverage result of the ML approaches is likely not caused by underestimating the uncertainty but by failing to capture the correct predictions for $z$. 

\begin{figure*}[ht]
    \centering
    \includegraphics[width=0.32\textwidth]{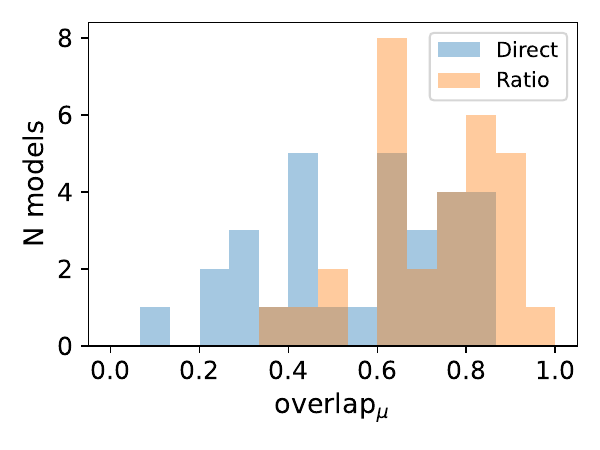}
    \includegraphics[width=0.32\textwidth]{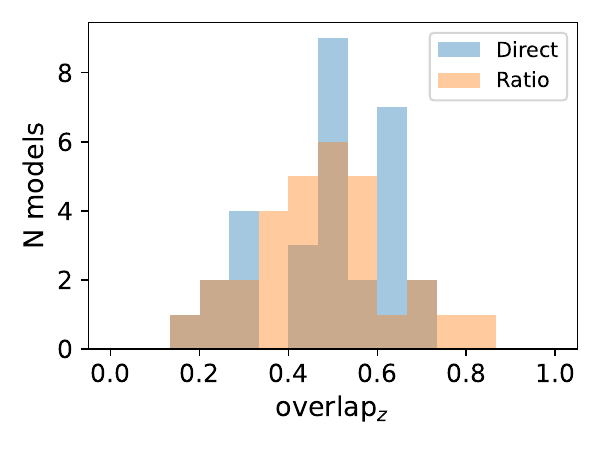}
    \includegraphics[width=0.32\textwidth]{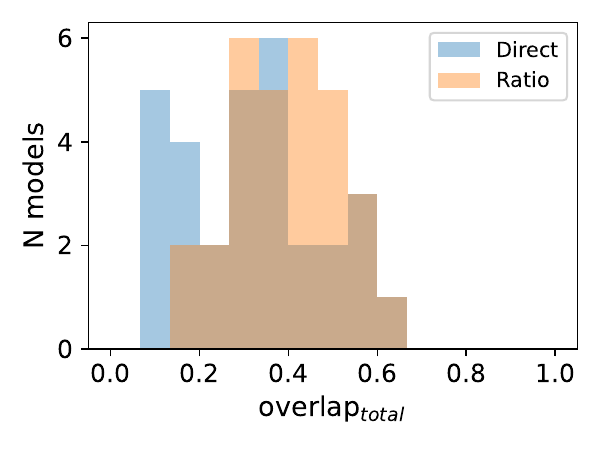}
    \includegraphics[width=0.34\textwidth]{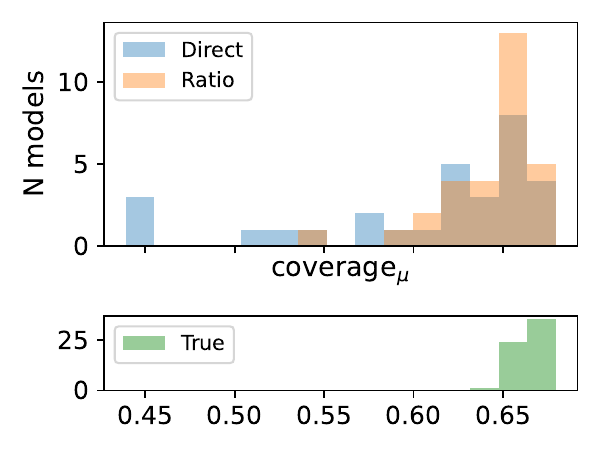}
    \includegraphics[width=0.34\textwidth]{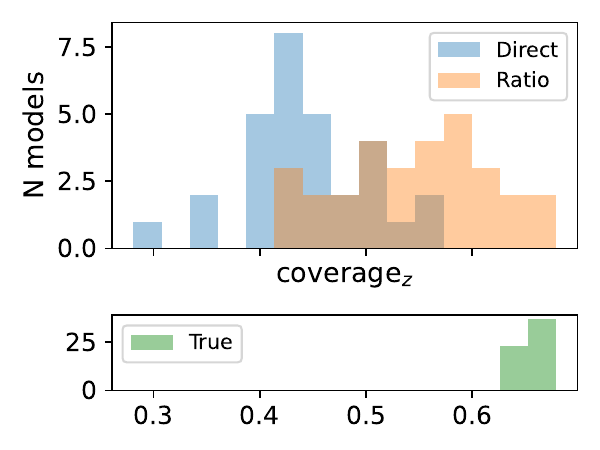}
    \includegraphics[width=0.34\textwidth]{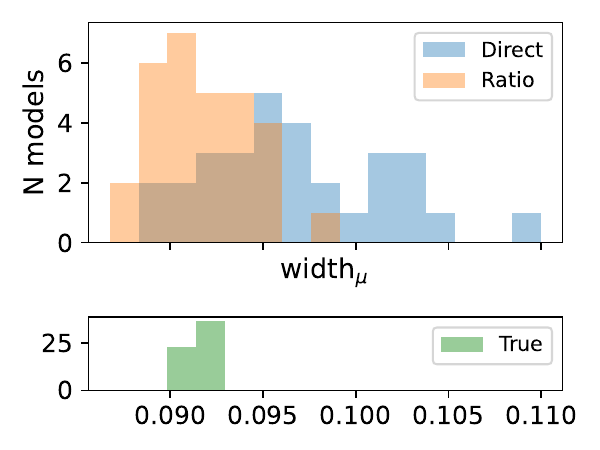}
    \includegraphics[width=0.34\textwidth]{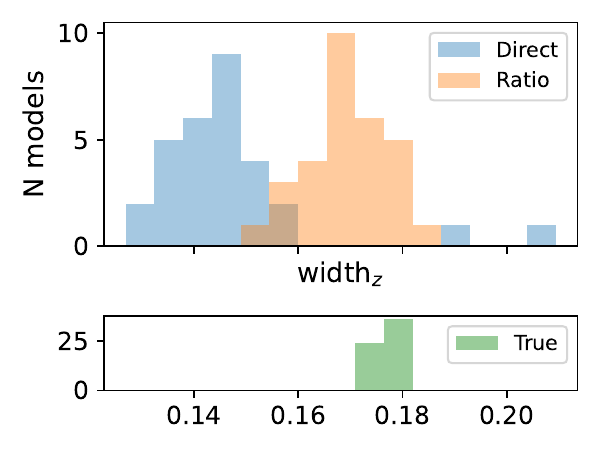}
    \caption{Gaussian small distance case results for the \flow (learning rate = $10^{-5}$) and \classi model (learning rate = $3\times 10^{-5}$). Histograms show the performance of 30 independently trained models of both the \classi model and the \flow model. For the coverage and mean width results, the plots are split into an upper section for the model results and a lower section for the ground truth model, indicating approximately what the histograms should look like if all 30 models were the perfect model.}
    \label{fig:hist_small_dist}
\end{figure*}

Figure~\ref{fig:hist_small_dist} shows the same evaluation performed on the small distance Gaussian dataset. 
Notably, the small distance between the two Gaussian peaks makes the task of classifying signal and background significantly more challenging than in the large distance case. 
This effect can clearly be seen in both the $\mu$ overlap and $\mu$ coverage, both of which show significantly larger deviations from the correct values than what was observed in the small distance case.
The smaller distance between peaks also makes it more challenging to determine their relative rotation $z$, which is visible in the $z$ coverage. 
Here, we also see a notable difference between the \flow model and \classi model. 
While both models have imperfect coverage, the \flow model performs notably worse, having an average coverage of only around $0.45$.
Looking at the width, we can further see that the \flow model predicts a $z$ width significantly below the width derived from the true likelihood. 
This indicates that the \flow model systematically underestimates the uncertainty, resulting in the incorrect coverage.
Importantly, this difference between the $z$ performance of the \flow and \classi models is not apparent from the overlap and can only be detected with the coverage metric. 

\begin{figure}[hbt]
    \centering
    \includegraphics[width=0.49\linewidth]{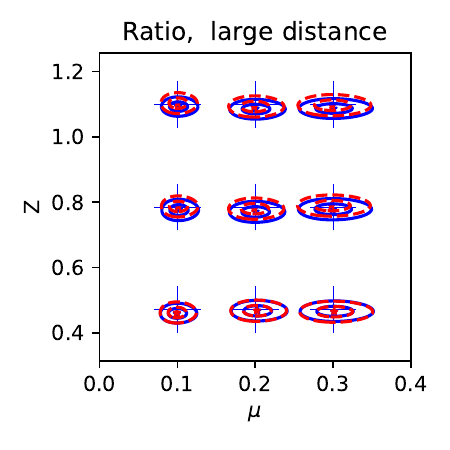}
    \includegraphics[width=0.49\linewidth]{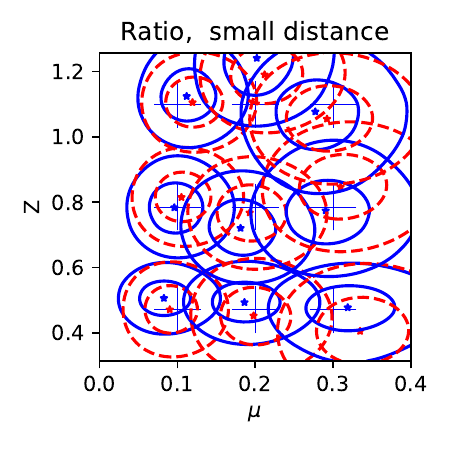}
    \includegraphics[width=0.49\linewidth]{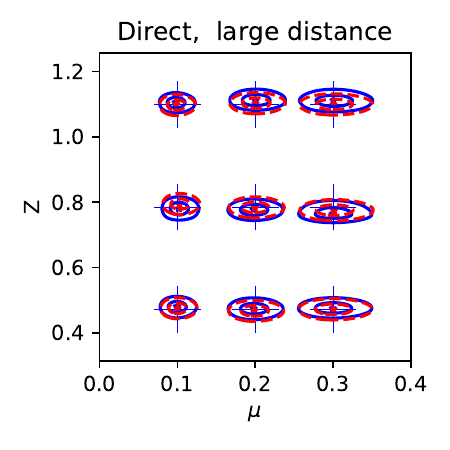}
    \includegraphics[width=0.49\linewidth]{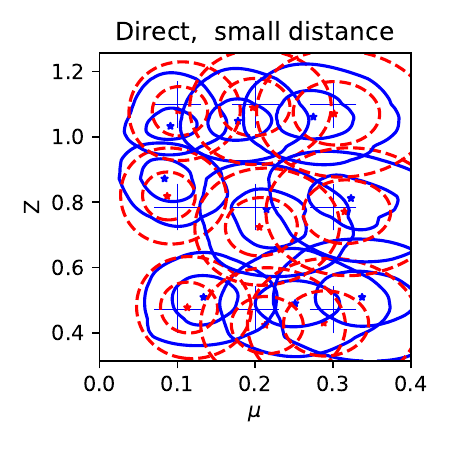}
    \caption{Examples of contours in the Gaussian case. 
    The upper row shows results for the \classi approach, while the lower row shows results for the \flow approach. The results in the left column were obtained using the large distance Gaussian data, and the results using the right column were trained using the small distance Gaussian data. 
    Metric averages over the test sets in each figure can be found in the Appendix in Table~\ref{tab:cont_gauss}.
    %\textbf{Normalizing flow, large distance case}: $\mu$ overlap=0.95, $z$ overlap=0.47, total overlap=0.47, $\mu$ coverage=0.65, $z$ coverage=0.52, $\mu$ mean width=0.035, $z$ mean width=0.032\\
    %\textbf{Classifier, large distance case}: $\mu$ overlap=0.94, $z$ overlap=0.52, total overlap=0.52, $\mu$ coverage=0.66, $z$ coverage=0.60, $\mu$ mean width=0.036, $z$ mean width=0.031\\
    %\textbf{Normalizing flow, small distance case}: $\mu$ overlap=0.55, $z$ overlap=0.41, total overlap=0.29, $\mu$ coverage=0.61, $z$ coverage=0.50, $\mu$ mean width=0.096, $z$ mean width=0.15\\
    %\textbf{Classifier, small distance case}: $\mu$ overlap=0.69, $z$ overlap=0.47, total overlap=0.41, $\mu$ coverage=0.64, $z$ coverage=0.63, $\mu$ mean width=0.089, $z$ mean width=0.17. \bpn{This caption is very ugly}
    }
    \label{fig:cont_gauss}
\end{figure}

Lastly, we can also examine the contours of the estimated likelihood in comparison to the true values for various combinations of $\mu$ and $z$ as shown in Fig.~\ref{fig:cont_gauss}. 
There is good agreement between the likelihood contours for the large distance case, while the small distance data set shows notable differences between model prediction and truth.
This is in line with the observations from the previous figures. 
Overall, both the \flow model and the \classi model perform comparably on the large distance case. 
For the more challenging small distance Gaussians, the \classi model has the same coverage but better width compared to the \flow model in $\mu$, and displays a better coverage than the \flow model in $z$. 
While only a single set of examples (with a single set of hyperparameters), this leads to the hypothesis that the \classi approach is more sensitive than the \flow approach.  This hypothesis will be tested in the collider physics example in the next section.
%
%This indicates that for a more challenging problem, the ratio-based model can achieve a good performance more easily compared to the flow model. }
%\bpn{Some statement about which is better or not.}
%

\subsection{Higgs Example}
\begin{figure}[hbt]
    \centering
    \includegraphics[width=0.49\linewidth]{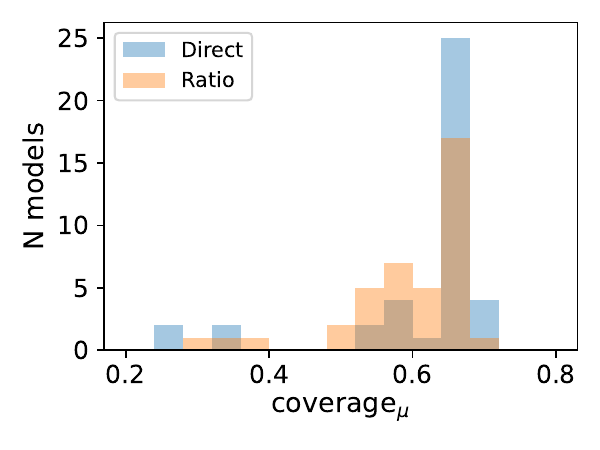}
    \includegraphics[width=0.49\linewidth]{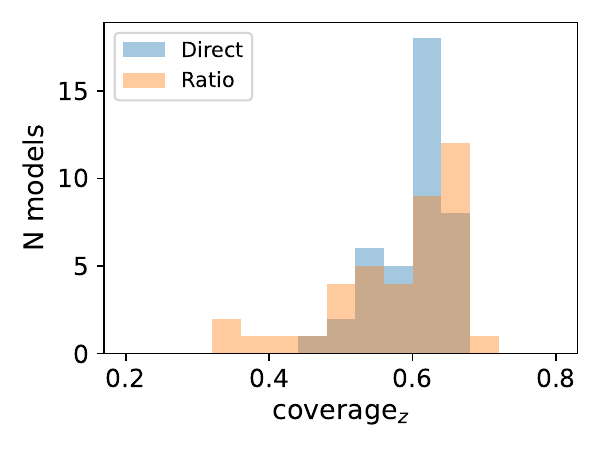}
    \includegraphics[width=0.49\linewidth]{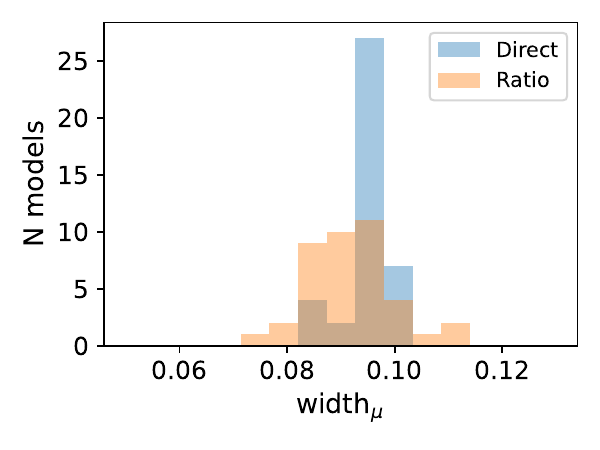}
    \includegraphics[width=0.49\linewidth]{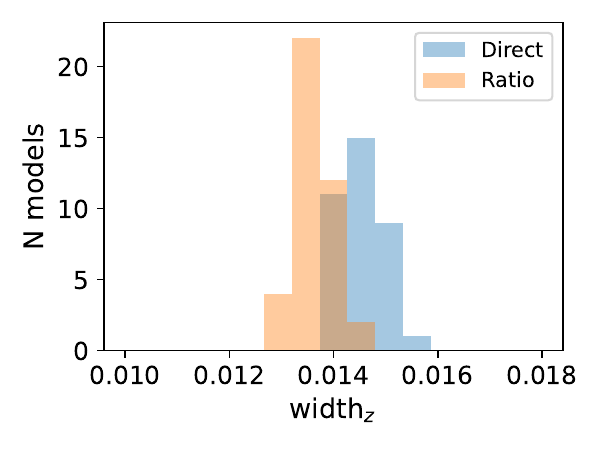}
    \caption{Higgs case results. Histograms show the average performance of 40 independently trained models of both the \classi model and \flow model.}
    \label{fig:hist_higgs}
\end{figure}

\begin{figure}[hbt]
    \centering
    \includegraphics[width=0.49\linewidth]{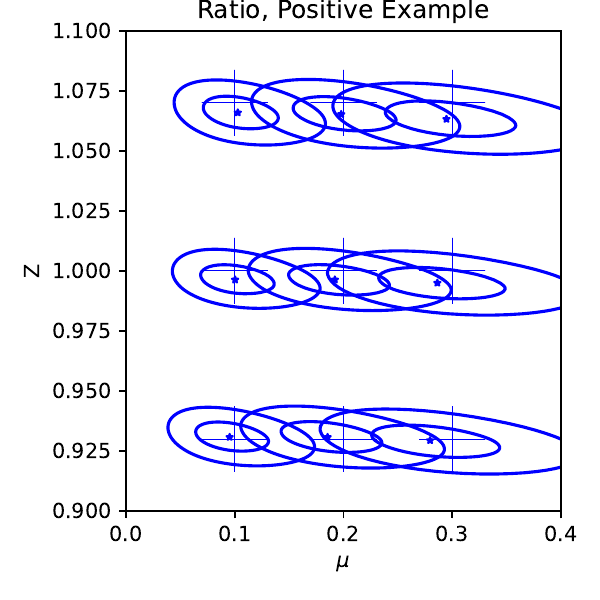}
    \includegraphics[width=0.49\linewidth]{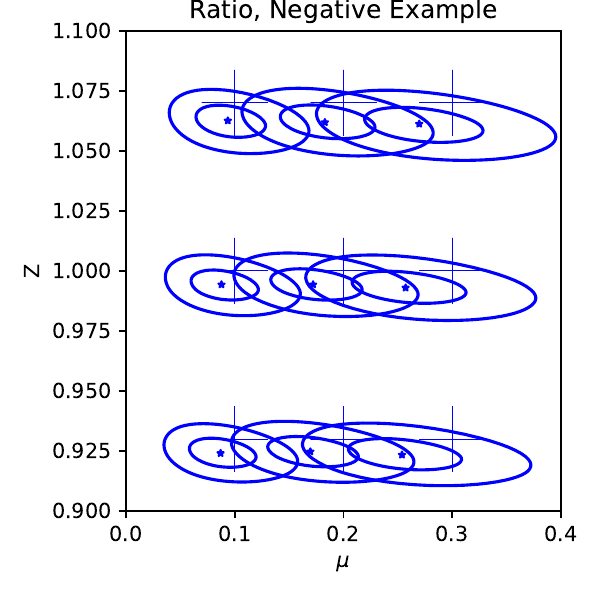}
    \includegraphics[width=0.49\linewidth]{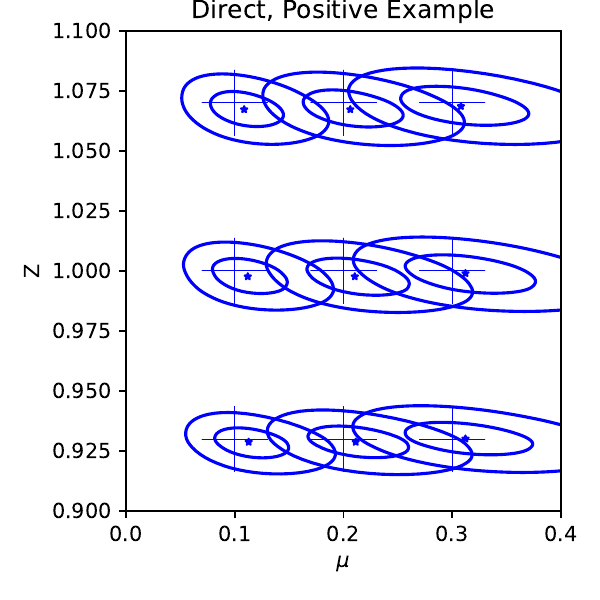}
    \includegraphics[width=0.49\linewidth]{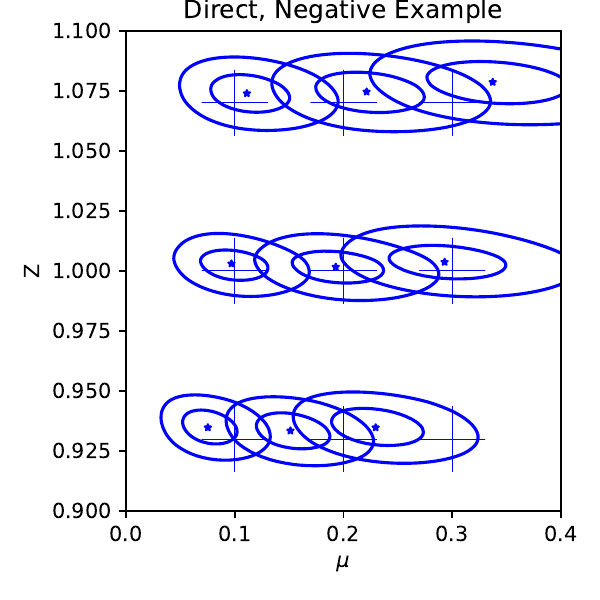}
    \caption{Examples of contours for models trained on the Higgs data. \textit{Positive Example} shows well-performing models, while \textit{Negative Example} shows models with lower performance. 
    Metric averages over the test sets in each figure can be found in the Appendix in Table~\ref{tab:cont_higgs}.
    }
    \label{fig:cont_higgs}
\end{figure}

% \begin{table}[h!]
%     \begin{tabular}{ |c|c|c| }
%         \hline
%                    & flow         & classifier   \\
%         \hline
%         coverage   & ...   & ...   \\
%         \hline
%         width & ... & ... \\
%         \hline
%     \end{tabular}
% \caption{Coverages and width results for flow and classifier. Results were taken for a range of learning rates and values shown here are for the learning rates with highest mean coverage ($10^{-5}$ for flow and $3\cdot10^{-4}$ for classifier).}
% \end{table}
%
We perform an identical analysis with the Higgs data, as was done with the Gaussian data, except that the number of trained models was increased to $N=40$.
Fig.~\ref{fig:hist_higgs} shows the same results that were discussed in the Gaussian case, except without any overlap metrics or the true likelihood results, as these would require access to the underlying likelihood of the Higgs data, which is not available. 
Nevertheless, the coverage metric results shown in the upper row show that both models have a $\mu$ coverage peaked around $0.675$, and a $z$ coverage peaked around $0.625$, which correspond to the correct coverage for $\mu$, and a slightly suboptimal coverage for $z$. 
It should be noted, however, that without access to true likelihood results, it is not possible to disentangle whether the $z$ coverage results are purely caused by poor model performance, or by an effect of the bootstrapping. 
Further, the width plots on the bottom row allow us to differentiate the performance of \classi and \flow models, where the \flow model has an, on average, larger width than the \classi model in both $\mu$ and $z$, even though the classifier displays a larger spread in $\mu$ width for different model trainings.  These findings are consistent with the hypothesis from the Gaussian case, and it would be interesting to see how other examples compare with these two.
Fig~\ref{fig:cont_higgs} shows the contours for individual models. Since we do not have a true contour as a benchmark, we instead illustrate two examples.
\textit{Positive Example} shows models with good coverage scores, while \textit{Negative Example} shows the contours of models with less optimal coverages. 
From this, we can see that there appears to be a correlation between the coverage score and the ability of a model to determine the correct values for $\mu$ and $z$.
This further demonstrates the usefulness of the coverage as a performance metric for uncertainty-aware models. 

\section{Conclusions and Outlook}
\label{sec:conclusions}

%Talk about calibration and other stategies for fixing the problems we find.

We have directly compared direct likelihood ratio estimation and likelihood ratio estimation for neural simulation-based parameter inference.  Our motivation for this study was the growing number of studies focusing on only one of these two approaches.  We found that both methods were able to effectively extract parameters with uncertainties.  On the specific examples we studied and with the hyperparameters selected, the likelihood-ratio method was more accurate and/or precise. However,  we found that the results are quite sensitive to the setup, including the data, network architecture/training, and inference evaluation protocol.  Extensive scans are computationally expensive, making even the basic task of hyperparameter optimization challenging.  While these issues make the comparison between approaches difficult, they also highlight the overall difficulty of optimizing an NSBI setup.  It will be interesting to see if the trends observed in this paper persist with higher-dimensional feature and parameter spaces and with advanced regularization techniques including ensembling, gradient-based parameter inference, and calibration~\cite{cranmer}.

\clearpage

\section*{Code and Data}

The software for this paper can be found on \href{https://github.com/SaschaDief/generative_vs_classifier} {Github}. A slightly updated version of the Higgs dataset is available at \href{https://www.codabench.org/competitions/2164/}{this Codabench competition}, and the most recent version of the uncertainty-aware Higgs classification challenge data set can be found \href{https://www.codabench.org/competitions/2977/}{here}.  

\section*{Acknowledgments}

BN and SD are supported by the U.S. Department of Energy (DOE), Office of Science under contract DE-AC02-05CH11231.  BS is supported by the Leiden University International Study Fund.  We thank Dorothea Samtleben for detailed feedback on the study and for many useful discussions with the FAIR Universe team, including Ragansu Chakkappai, Po-Wen Chang, Yuan-Tang Chou, Jordan Dudley, Steven Farrell1, Aishik Ghosh, Isabelle Guyon, Chris Harris, Shih-Chieh Hsu, Elham E Khoda, David Rousseau, Ihsan Ullah, and Yulei Zhang.

\bibliography{main}
\bibliographystyle{JHEP}

%\newpage

\appendix
\section{Numerical Results of Example Contours}

In this Appendix, we present the coverage, mean width, and overlap values for the models that correspond to the contours shown in Fig.~\ref{fig:cont_gauss} and Fig.~\ref{fig:cont_higgs}. The reported values are averaged over the 9 test sets shown in the contour plots. 
\begin{table}[hbt]
    \centering
    \begin{tabular}{c|c|c|c|c}
     Gaussian & \multicolumn{2}{c}{large distance}  & \multicolumn{2}{c}{small distance} \\
        \hline
         & \classi & \flow & \classi  & \flow \\
        \hline
        overlap$_{\mu}$          & 0.94 & 0.95 & 0.69 & 0.55 \\
        overlap$_{z}$            & 0.52 & 0.47 & 0.47 & 0.41 \\
        overlap$_{\textrm{tot}}$ & 0.52 & 0.47 & 0.41 & 0.29 \\
        coverage$_{\mu}$         & 0.66 & 0.65 & 0.64 & 0.61 \\
        coverage$_{z}$           & 0.60 & 0.52 & 0.63 & 0.50 \\
        width$_{\mu}$            & 0.036& 0.035& 0.089& 0.096\\
        width$_{z}$              & 0.031& 0.032& 0.17 & 0.15 \\
    \end{tabular}
    \caption{Coverage, mean width and overlap values of the models on the Gaussian data, for which the contours were shown in Fig.~\ref{fig:cont_gauss}. The table is separated into the large distance case and the small distance case. Further, the table is subdivided into the \classi and \flow approaches.}
    \label{tab:cont_gauss}
\end{table}
%
%\textbf{Classifier, small distance case}: $\mu$ overlap=0.69, $z$ overlap=0.47, total overlap=0.41, $\mu$ coverage=0.64, $z$ coverage=0.63, $\mu$ mean width=0.089, $z$ mean width=0.17. \\}
%\textbf{Normalizing flow, small distance case}: $\mu$ overlap=0.55, $z$ overlap=0.41, total overlap=0.29, $\mu$ coverage=0.61, $z$ coverage=0.50, $\mu$ mean width=0.096, $z$ mean width=0.15\\
%
%\textbf{Classifier, large distance case}: $\mu$ overlap=0.94, $z$ overlap=0.52, total overlap=0.52, $\mu$ coverage=0.66, $z$ coverage=0.60, $\mu$ mean width=0.036, $z$ mean width=0.031\\
%\textbf{Normalizing flow, large distance case}: $\mu$ overlap=0.95, $z$ overlap=0.47, total overlap=0.47, $\mu$ coverage=0.65, $z$ coverage=0.52, $\mu$ mean width=0.035, $z$ mean width=0.032\\
%
\begin{table}[hbt]
    \centering
    \begin{tabular}{c|c|c|c|c}
        Higgs data &  \multicolumn{2}{c}{high agreement} & \multicolumn{2}{c}{low agreement} \\
        \hline
         & \classi & \flow & \classi & \flow \\
        \hline
        coverage$_{\mu}$ & 0.67 & 0.68 & 0.60 & 0.56 \\
        coverage$_{z}$   & 0.68 & 0.66 & 0.48 & 0.59 \\
        width$_{\mu}$    & 0.098 & 0.095 & 0.085 & 0.087\\
        width$_{z}$      & 0.014 & 0.014 & 0.013 & 0.014\\
    \end{tabular}
    \caption{Coverage and mean width values of the models on the Higgs data, for which the contours were shown in Fig.~\ref{fig:cont_higgs}. The table is separated into the models that agree well with the truth (left) and the instance where the models agree less well (right). Further, the table is subdivided into the \classi and \flow approaches.}    
    \label{tab:cont_higgs}
\end{table} 
%
%Metric averages over the 9 testsets (all drawn from the same dataset) in each figure are: 
%\textbf{Flow, example 1} $\mu$ coverage=0.68, $z$ coverage=0.66, $\mu$ mean width=0.095, $z$ mean width=0.014. 
%\textbf{Flow, example 2} $\mu$ coverage=0.56, $z$ coverage=0.59, $\mu$ mean width=0.087, $z$ mean width=0.014. 
%\textbf{Class, example 1} $\mu$ coverage=0.67, $z$ coverage=0.68, $\mu$ mean width=0.098, $z$ mean width=0.014. %
%\textbf{Class, example 2} $\mu$ coverage=0.60 and $z$ coverage=0.48, $\mu$ mean width=0.085, $z$ mean width=0.013.

\end{document}